\documentclass[aps,prc,groupedaddress,showpacs,preprint]{revtex4-1}
\usepackage{graphicx}
\usepackage{colordvi}

\begin{document}

\title{Constraining the high-density nuclear symmetry energy with the transverse-momentum
dependent elliptic flow}
\author {Yongjia Wang$\, ^{1,2}$,
Chenchen Guo$\, ^{1,3}$,
Qingfeng Li$\, ^{1}$\footnote{E-mail address: liqf@hutc.zj.cn},
Hongfei Zhang$\, ^{2}$,
Y. Leifels$\, ^{4}$,
and
W. Trautmann$\, ^{4}$}

\affiliation{
1) School of Science, Huzhou Teachers College, Huzhou 313000, P.R. China \\
2) School of Nuclear Science and Technology, Lanzhou University, Lanzhou 730000, P.R. China \\
3) College of Nuclear Science and Technology, Beijing Normal University, Beijing 100875, P.R. China \\
4) GSI Helmholtzzentrum f\"ur Schwerionenforschung GmbH, D-64291 Darmstadt, Germany\\
\\
 }
\date{\today}

\begin{abstract}
Within the newly updated version of the ultrarelativistic quantum molecular dynamics (UrQMD)
model, the transverse-velocity dependence of the elliptic flow of free nucleons
from $^{197}$Au+$^{197}$Au collisions at the incident energy 400 MeV$/$nucleon is studied
within different windows of the normalized c.m. rapidity $y_0$.
It is found that the elliptic flow difference
$v_{2}^{n}$-$v_{2}^{p}$ and ratio $v_{2}^{n}$/$v_{2}^{p}$ of neutrons versus protons are
sensitive to the density dependence of the symmetry energy, especially the ratio
$v_{2}^{n}$/$v_{2}^{p}$ at small transverse velocity in the intermediate rapidity
intervals $0.4<|y_0|<0.6$.
By comparing either transverse-momentum dependent or integrated FOPI/LAND elliptic flow data
of nucleons and hydrogen isotopes with
calculations using various Skyrme interactions, all exhibiting similar values of isoscalar incompressibility
but very different density dependences of the symmetry energy, a moderately soft to linear
symmetry energy is extracted, in good agreement with previous UrQMD or T\"{u}bingen
QMD model calculations but contrasting results obtained with $\pi^-/\pi^+$ yield ratios
available in the literature.
\end{abstract}


\maketitle

The nuclear symmetry energy, which plays an important role in studying exotic nuclei,
heavy-ion collisions (HICs) with and without radioactive beams, and neutron stars,
is a very active field of research in nuclear and astronuclear physics.
It can be calculated from the parabolic approximation of the equation of state (EoS)
of isospin asymmetric nuclear matter,
$e(\rho,\delta)=e_{0}(\rho,0)+e_{\rm sym}(\rho)\delta^{2}$,
where $\delta=(\rho_{n}-\rho_{p})/(\rho_{n}+\rho_{p})$ is the
isospin asymmetry defined through the neutron ($\rho_n$) and proton
($\rho_p$) densities.
The first term $e_{0}(\rho,0)$ is the energy per nucleon of isospin symmetric nuclear matter
which, at the saturation density $\rho_0$, is known as the binding energy $E_0$ of nuclear matter;
the coefficient $e_{\rm sym}(\rho)$ of the second term is the bulk symmetry energy.
There exists a large number of theoretical predictions for the density dependence of
$e_{\rm sym}$ obtained with different many-body, effective field, or phenomenological
approaches. For a recent review, see Ref.~\cite{BALi08}.

Many practical attempts have been made to estimate parameters of the symmetry energy around
saturation. They include the symmetry energy coefficient $S_0=e_{\rm sym}(\rho_{0})$,
the slope parameter
$L=3\rho_{0}\left(\frac{\partial{e_{\rm sym}(\rho)}}{\partial\rho}\right)|_{\rho=\rho_{0}}$,
and the curvature parameter
$K_{\rm sym}=9\rho_{0}^2\left(\frac{\partial^2{e_{\rm sym}(\rho)}}{\partial\rho^2}\right)|_{\rho=\rho_{0}}$,
and are obtained from the comparison of model calculations and experimental data, such as
atomic masses~\cite{Agrawal:2013hha,Liu:2010ne,Chen:2011ek,moller},
$\alpha$-decay energies~\cite{Dong:2012ah},
nuclear charge radii~\cite{Wang:2013zia},
the thickness of the neutron skins of heavy nuclei~\cite{Centelles:2008vu,Chen:2010qx},
pygmy and giant dipole resonances~\cite{Carbone:2010az,Trippa:2008gr,RocaMaza:2012mh},
and nuclear isobaric analog-state energies~\cite{Danielewicz:2008cm}.
Even though the deduced constraints are somewhat different in different studies, they are
generally consistent with each other~\cite{Li:2012mw,Chen:2012pk,Lattimer:2012xj,Tsang:2012se}.
Examples are the values $S_0=31\pm2$~MeV and $L=50\pm20$~MeV reported in
Refs.~\cite{Li:2012mw,Chen:2012pk} and the results of Refs.~\cite{Tsang:2012se,Lattimer:2012xj}
providing a constraint centered around ($S_0$, $L$) = (32.5, 70) MeV.

To probe the high-density behavior of the nuclear symmetry energy in terrestrial laboratories,
we need the aid of HICs. Usually,
the measured experimental data are compared with
corresponding results of microscopic transport models in order to extract the information they
carry with regard to properties of the symmetry energy.
Several observables have been found or predicted
to be sensitive to the nuclear symmetry energy as, e.g., neutron and proton yields and flow
ratios, double ratios, or differences, $\pi^{-}$/ $\pi^{+}$ and $K^0/K^+$ meson production
ratios, the
$\Sigma^{-}$/$\Sigma^{+}$ ratio, and the balance energy of directed flow~\cite{DiToro:2010ku,
Tsang:2008fd,Lopez:2007,Xiao:2009zza,Feng:2009am,Russotto:2011hq,Cozma:2011nr,Xie:2013np,Cozma:2013sja,
Li:2005zza,LI:2005zi,Kumar:2011td,Gautam:2010da,Wang:2012sy,Guo:2012aa}. However,
even though precise experimental data are available for some of
these quantities, their interpretation is strongly model dependent and the obtained
constraints on the nuclear symmetry energy at high densities
are not consistent with each other
(see, e.g., Refs.~\cite{Tsang:2012se,Li:2012mw,Chen:2012pk,Wolter}).

Recently, the neutron-proton (or neutron-hydrogen) elliptic flow difference $v_2^n-v_2^{p,H}$ and
ratio $v_2^n/v_2^{p,H}$ have been taken to constrain the high-density behavior of the nuclear
symmetry energy. The elliptic flow parameter $v_2$ is the second-order coefficient in the
Fourier expansion of the azimuthal distribution of detected particles
\begin{equation}
\frac{dN}{d\phi}=v_0\left[ 1+2v_1\cos \left ( \phi \right ) +2v_2  \cos \left (  2 \phi \right ) \right]
\label{azieq}
\end{equation}
and has the properties
\begin{equation}
v_2 \equiv \left \langle \cos \left (  2 \phi \right ) \right \rangle =\frac{p_x^2-p_y^2}{p_t^2}.
\label{v2eq}
\end{equation}
Here $\phi$ denotes the azimuthal angle of the considered outgoing particle
with respect to the reaction plane and $p_x$ and $p_y$ are the two components of the transverse
momentum $p_t=\sqrt{p_x^2+p_y^2}$ in the so oriented frame. The angular bracket denotes an
average over all considered particles of a given event class. The parameter $v_2$ is thus a
function of particle type, impact parameter, rapidity $y$, and transverse momentum $p_t$.
By comparing calculations of two different versions of quantum molecular dynamics (QMD) models
with the existing FOPI/LAND experimental data, a moderately soft to linear symmetry energy with
a density dependence of its potential term proportional to $(\rho/\rho_{0})^{\gamma}$ with the
strength parameter $\gamma$=0.9$\pm$0.4 (correspondingly, $L=83\pm26$ MeV) and a moderately
stiff to linear symmetry energy with the stiffness parameter $x =-1.35\pm 1.0$ of the generalized
Gogny force~\cite{das03} (correspondingly, $L=122\pm57$ MeV), have been reported in
Refs.~\cite{Russotto:2011hq} and~\cite{Cozma:2013sja}, respectively.

Furthermore, by using the newly updated ultrarelativistic quantum molecular dynamics (UrQMD)
model in which the Skyrme potential energy density functional is introduced, the recently
published flow data~\cite{FOPI:2010aa,FOPI:2011aa} of the FOPI Collaboration for light charged
particles (protons, $^2$H, $^3$H, $^3$He, and $^4$He) can be reproduced quite
well~\cite{Wang:2013wca}.
An advantage of the UrQMD update is that the stiffness of the symmetry energy can be more
consistently selected within a broad range by simply
changing Skyrme interactions, rather than by varying the exponent $\gamma$ in the potential term
of the symmetry energy which, in addition, cannot be used to express a very soft symmetry
energy~\cite{Dong:2012zza}. It thus seems worthwhile to update the results of
Ref.~\cite{Russotto:2011hq} within the framework of the newly updated UrQMD model,
utilizing different Skyrme interactions with similar values of the isoscalar incompressibility
but differing strengths of the symmetry energy.

In the work reported in Ref.~\cite{Cozma:2013sja}, performed with a version of the T\"{u}bingen QMD
model, the effects of uncertainties in several important ingredients of the transport model on
the neutron-proton elliptic flow difference and ratio were carefully analyzed.
They include, in addition to the symmetry energy, the incompressibility $K_0$ of nuclear matter,
the nuclear optical potential, the in-medium nucleon-nucleon cross sections, and
the momentum dependence of the symmetry potential.
This work is important since most of these ingredients are still largely uncertain
and their effects need to be assessed.
In the current work an alternative strategy is pursued, consisting of firstly finding a set of
model parameters which can ''best'' describe the FOPI flow data for charged light clusters,
then exploring the sensitivity of the symmetry energy to the neutron-proton flow difference
and ratio in various rapidity and transverse momentum intervals,
and finally estimating the value of the slope parameter
of the symmetry energy by comparing with the existing data. The overall model dependence can
still be discussed during this process
and will become apparent when comparing the calculations with the two members of the QMD family.

The UrQMD model has been widely and successfully used to study nuclear reactions of \emph{p}+\emph{p}, \emph{p}+\emph{A} and \emph{A}+\emph{A} systems within a
large range of beam energies, from SIS up to the LHC~\cite{Bass98,Bleicher:1999xi,Li:2011zzp,Li:2012ta}. In the present code~\cite{Wang:2013wca}, the nuclear effective interaction potential energy $U$ of the Hamiltonian $H$ is derived from the integration of the Skyrme potential energy density functional, $U_{\rho}=\int u_{\rho}d^3\vec{r}$, and $u_{\rho}$ reads
\begin{eqnarray}
u_{\rho}=\frac{\alpha}{2}\frac{\rho^2}{\rho_0}+
\frac{\beta}{\eta+1}\frac{\rho^{\eta+1}}{\rho_0^{\eta}}+
\frac{g_{sur}}{2\rho_0}(\nabla\rho)^2
+\frac{g_{sur,iso}}{2\rho_0}[\nabla(\rho_n-\rho_p)]^2\nonumber\\
+(A\rho^{2}+B\rho^{\eta+1}+C\rho^{8/3})\delta^2
+g_{\rho\tau}\frac{\rho^{8/3}}{\rho_0^{5/3}}  \label{urho}
\end{eqnarray}
where $\alpha$, $\beta$, $\eta$, $g_{sur}$, $g_{sur,iso}$, \emph{A}, \emph{B},\emph{ C},
and $g_{\rho\tau}$ are parameters which can be directly calculated by using Skyrme parameters
(see, e.g., Refs.~\cite{Wang:2013wca,Zhang:2006vb}).
It is known that the $\rho\tau$ term, obtained from the Thomas-Fermi approximation to the
kinetic energy density, can not fully represent the momentum dependence of the whole
non-equilibrium dynamic process~\cite{Zhang:2006vb}.
The momentum dependence of the real part of the optical potential (called ''optical potential''
for short in this paper) originating from that in Ref.~\cite{Aichelin:1987ti} is also considered,
as well as the  Coulomb term. The importance of the optical potential for observables such as
particle production and flow measured in HICs at intermediate energies has been widely
investigated but its form is still far from being settled~\cite{Aichelin:1987ti,Hartnack:1994zz}.
Recently, an isospin-dependent optical
potential~\cite{das03} was introduced into  the isospin-dependent Boltzmann-Uehling-Uhlenbeck
(IBUU04) transport model based on the Gogny effective interactions~\cite{Li:2003zg}, giving rise to the issues of the momentum dependence of the symmetry energy and the associated neutron-proton effective mass splitting effect in the isospin asymmetric nuclear matter. Effects of the momentum-dependent symmetry potential or the neutron-proton effective mass splitting in HICs have been discussed recently \cite{Li:2003zg,Chen:2004kj,Baran:2004ih,Giordano:2010pv,Feng:2011pu,Gao:2011sg,Zhang:2012fc,Feng:2011xp}. But, it is still unclear whether the effective mass of neutrons is larger or smaller than that of protons in, for example, the neutron-rich nuclear medium \cite{BALi08,DiToro:2005ac,vanDalen:2005ns,Zhang:2014sva}. While it is claimed that
the neutron-proton elliptic flow difference is influenced to some extent by the momentum dependence of the nuclear symmetry potential in Refs.~\cite{Giordano:2010pv,Feng:2011xp}, it was reported in Ref.~\cite{Zhang:2012fc} that the flow difference does not exhibit a visible sensitivity to the momentum dependent part of the isovector nucleon potential within the constraint of an asy-soft EoS, although the neutron-proton differential transverse flow does. Therefore, the influence of the neutron-proton effective mass splitting on observables in HICs certainly deserves further studies. However, together with support from previous QMD analyses, an isospin-independent form of the optical potential is still considered as appropriate for the current analysis of elliptic flows.

\begin{table}[htbp]
\caption{\label{tab:table1} Saturation properties of nuclear matter as obtained with
selected Skyrme parameterizations used in this work.}
\begin{ruledtabular}
\begin{tabular}{lccccccccccccc}
&$\rho_{0}(fm^{-3})$
&&$E_0(MeV)$
&&$K_0(MeV)$
&&$S_0(MeV)$
&&$L (MeV)$
&&$K_{\rm sym}(MeV)$
&&$m^{*}/m$ \\
\hline
Skz4 &0.160 &&-16.01 &&230.08 &&32.01&&5.75&&-240.86 &&0.70\\
Skz2 &0.160 &&-16.01&&230.07&&32.01&&16.81&&-259.66&&0.70\\
SV-mas08&0.160 &&-15.90&&233.13&&30.00&&40.15&&-172.38&&0.80\\
MSL0    &0.160&&-16.00&&230.00&&30.00&&60.00&&-99.33&&0.80\\
SkA &0.155&&15.99&&263.16&&32.91&&74.62&&-78.46&&0.61\\
SV-sym34 &0.159&&-15.97&&234.07&&34.00&&80.95&&-79.08&&0.90\\
Ska35s25 &0.158&&-16.14&&241.30&&36.98&&98.89&&-23.57&&0.99\\
SkI5 &0.156&&-15.85&&255.79&&36.64&&129.33&&159.57&&0.58\\
SkI1 &0.160&&-15.95&&242.75&&37.53&&161.05&&234.67&&0.69\\

\end{tabular}

\end{ruledtabular}\label{skyrme}
\end{table}

In this work, we choose 19 Skyrme interactions Skz4, BSk8, Skz2, BSk5, SkT6, SV-kap00, SV-mas08,
SLy230a, SLy5, SV-mas07, SV-sym32, MSL0, SkO', Sefm081, SV-sym34, Rs, Sefm074, Ska35s25, and
SkI1~\cite{Dutra:2012mb} which give quite similar values of $K_0$ (within about $230\pm10$ MeV)
but different $L$ and $K_{\rm sym}$ values (the saturation properties of selected typical forces
are shown in Table~\ref{skyrme}). Moreover, in order to examine the influence of $K_0$ on the
isospin-sensitive observables such as the elliptic flow ratio and difference,
the parameterization sets SkA and SkI5~\cite{Dutra:2012mb}, giving larger incompressibility
values than other sets, are also included.
We notice from Table~\ref{skyrme} and also from Ref.~\cite{Dutra:2012mb} that, due to the
momentum-dependence of the Skyrme potential itself, values of the effective mass
ratio $m^*/m$ at normal density of selected sets still spread in a large range from
about 0.6 up to 1.0 while the adopted optical potential provides an acceptable value of
$m^*/m = 0.75$ at the Fermi momentum.

\begin{figure}[htbp]
\centering
\includegraphics[angle=0,width=0.8\textwidth]{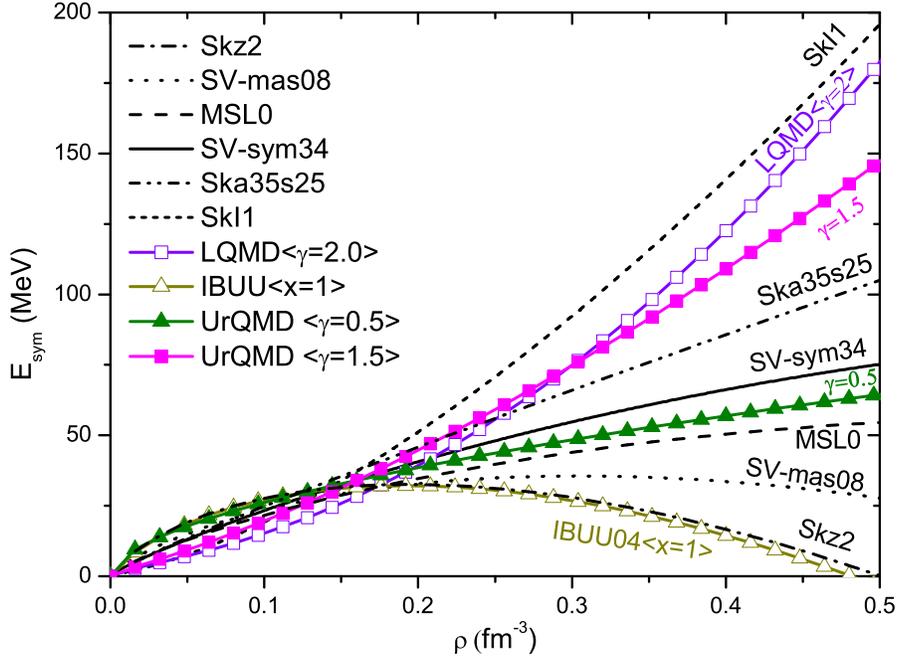}
\caption{\label{fig1}(Color online) Density dependence of the symmetry energy
for Skyrme interactions Skz2, SV-mas08, MSL0, SV-sym34, Ska35s25, and SkI1 (lines). Symmetry energies used in previous UrQMD~\cite{Russotto:2011hq} (lines with solid symbols), in LQMD~\cite{Feng:2009am} (line with open squares), and in IBUU04~\cite{Xiao:2009zza} (line with open triangles) calculations are also shown for comparison. }
\end{figure}

Figure~\ref{fig1} shows the density dependence of the symmetry energy for Skyrme interactions
Skz2, SV-mas08, MSL0, SV-sym34, Ska35s25, and SkI1.
In addition, the stiff (UrQMD-$\gamma=1.5$) and soft (UrQMD-$\gamma=0.5$)
symmetry energies used in previous UrQMD calculations are shown for comparison and
the symmetry energies deduced from the analyses of the FOPI $\pi^-/\pi^+$ ratios
within the IBUU04~\cite{Xiao:2009zza} and the Lanzhou quantum molecular
dynamics (LQMD)~\cite{Feng:2009am} models are exhibited as well.
It is apparent that the set of selected Skyrme forces covers the different forms of symmetry
energies presented and currently discussed by theoretical groups.

The treatment of the collision term is the same as in our previous work~\cite{Wang:2013wca}
in which the FP4 parameterization of the in-medium nucleon-nucleon
elastic cross section (NNECS) is adopted (except where stated otherwise).
The program is stopped at 150~fm$/$c and then
an isospin-dependent minimum span tree (iso-MST) algorithm is used to construct clusters.
Nucleons with relative distances smaller than $R_0$ and relative momenta smaller than $P_0$
are considered to belong to the same cluster. In the present work, $R_0$ and $P_0$ are set
to $R_0^{pp}=2.8$~fm, $R_0^{nn}=R_0^{np}=3.8$~fm, and $P_0 =0.25$~GeV/$c$.

\begin{figure}[htbp]
\centering
\includegraphics[angle=0,width=0.7\textwidth]{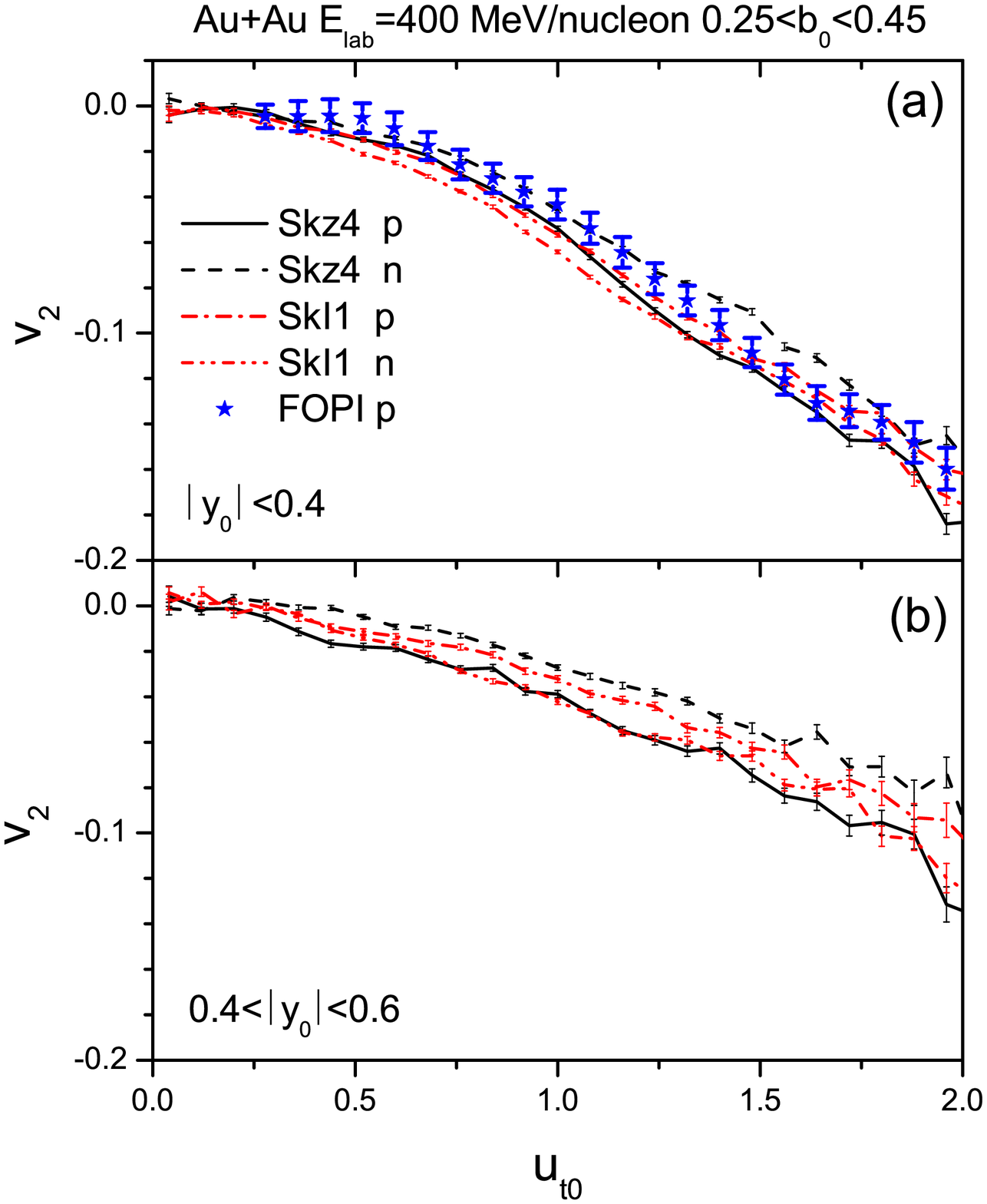}
\caption{\label{fig2}(Color online) Elliptic flow $v_2$ of protons and neutrons
in semi-central ($0.25<b_0<0.45$) $^{197}$Au+$^{197}$Au collisions at
$E_{\rm lab}=400$~MeV/nucleon as a function of the normalized transverse component $u_{t0}$
of the four-velocity.
Calculations with Skz4 and SkI1 (lines) for the two rapidity windows
$|y_0|<0.4$ (a) and $0.4<|y_0|<0.6$ (b) are shown and, in the top panel,
compared with the FOPI data for protons (stars) reported in Ref.~\protect\cite{FOPI:2011aa}.}
\end{figure}

As a first step, we try to describe the recent FOPI experimental data of the transverse
momentum dependence of elliptic flow~\cite{FOPI:2011aa} using the updated UrQMD model.
Figure~\ref{fig2} shows the $u_{t0}$ dependence of the elliptic flow of protons and
neutrons in semi-central ($0.25<b_0<0.45$) $^{197}$Au+$^{197}$Au collisions
at $E_{\rm lab}$=400 MeV/nucleon as calculated with the two parameter sets Skz4 and SkI1
for the two rapidity windows $|y_0|<0.4$ and $0.4<|y_0|<0.6$.
The reduced impact parameter $b_0$ is defined
as $b_0=b/b_{max}$ with $b_{max} = 1.15 (A_{P}^{1/3} + A_{T}^{1/3})$. The scaled units
$u_{t0}\equiv u_t/u_{p}$ and $y_0\equiv y/y_{p}$ are used as done in Ref.~\cite{FOPI:2011aa}.
The subscript $p$ refers to the incident projectile in the center-of-mass system, and the
subscript $t$ denotes transverse (spatial) components. Here $u_t=\beta_t\gamma$ is the
transverse component of the four-velocity $u=(\gamma, \bf{\beta} \gamma$) while the
longitudinal rapidity is $y=1/2ln((1+\beta_z)/(1-\beta_z))$.
The proton elliptic flow data for $|y_0|<0.4$~\cite{FOPI:2011aa} is shown with stars
while calculations are given by the lines.

Firstly, it is easily seen that the FOPI flow values of protons can be
reproduced fairly well by both parameter sets in the whole $u_{t0}$ region of the mid-rapidity
window. The difference of $K_0$ obtained with Skz4 and SkI1 is only about 12 MeV while that
of $L$ is more than 150 MeV (Table~\ref{skyrme}). This implies that
elliptic flows of nucleons can not be used individually to constrain the stiffness of the
symmetry energy although they are known to be sensitive to the isoscalar part of the EoS.
This is not unexpected as the contribution of the symmetry energy term is minor in
comparison.

Secondly, if one compares the medium-modified NNECS adopted here (FU3FP4), following the
results of Ref.~\cite{Wang:2013wca}, with the choice FU3FP1 suggested in Ref.~\cite{Li:2011zzp},
it is found that, in order to achieve the best description
of the excitation function of collective flows, the medium corrections of NNECS should be
gradually reduced with increasing beam energy. The restoration of the free NNECS should take place
at lower relative momenta at the higher energies, a trend also reported in Ref.~\cite{Zhang:2006vb}.
However, the need for an enhancement of the in-medium NNECS found there
when describing the elliptic flow of $Z\leq 2$ particles from midcentral
$^{197}$Au+$^{197}$Au collisions at about 400 MeV/nucleon is not evident here.
From the current analysis, we find that this is likely due to the different particle species
examined in the two studies. It is known that, mainly due to the lack of spin degrees of
freedom in the QMD-like models, the yield of $Z=2$ particles is largely underestimated.
When taking the contribution of light clusters weighted by their numbers into account,
irrespective of their mass numbers, the calculated absolute value of the elliptic flow
for $Z\leq 2$ particles becomes slightly underestimated,
an effect compensated with enhanced cross sections in the comparison with the experimental
data for $^{197}$Au+$^{197}$Au collisions reported in Ref.~\cite{Zhang:2006vb}.

Finally, for both rapidity windows, we observe a small but opposite effect of Skz4 and SkI1 on
proton and neutron elliptic flows. In the Skz4 case (soft symmetry energy), the squeeze-out
of protons is much larger than that of neutrons, whereas in the SkI1 case (stiff symmetry energy),
the trend is reversed: neutrons show a larger squeeze-out than protons. The differences grow with
increasing $u_{t0}$. As reported in Ref.~\cite{Russotto:2011hq}, this phenomenon reflects
the high density behavior of the symmetry energy.

\begin{figure}[htbp]
\centering
\includegraphics[angle=0,width=0.9\textwidth]{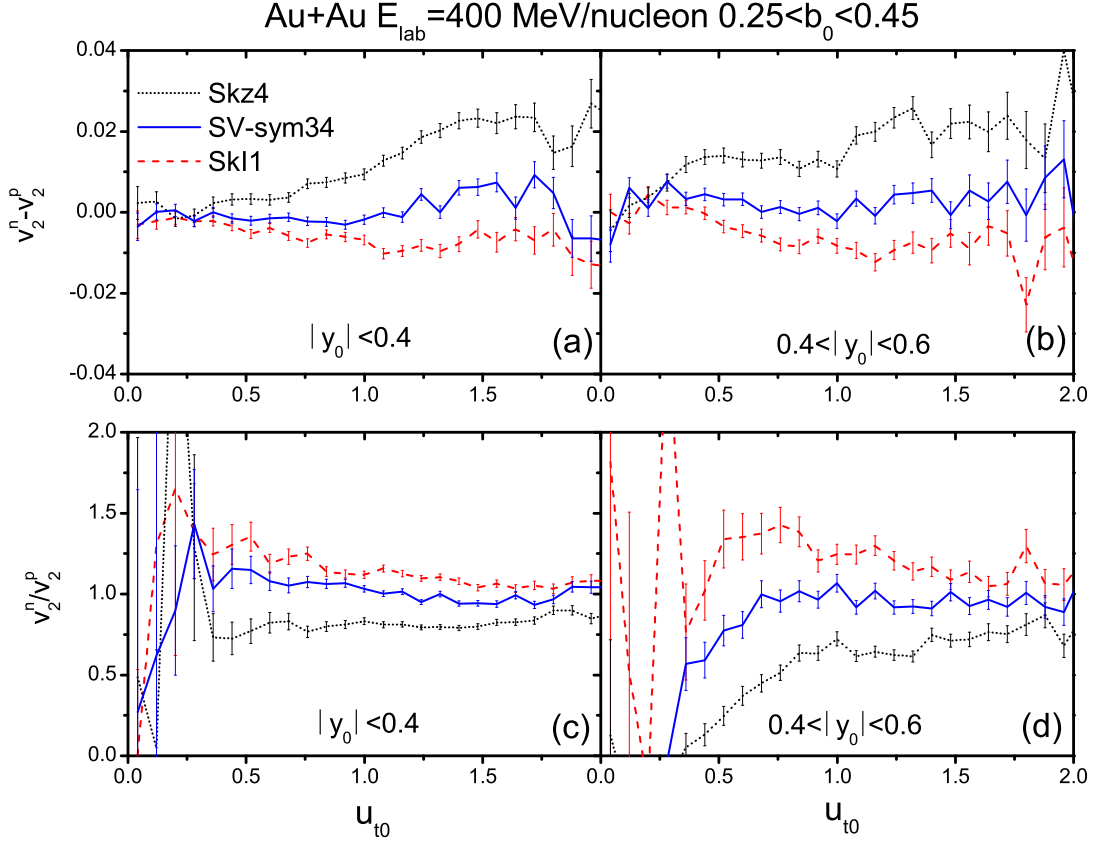}
\caption{\label{fig3}(Color online) Dependence of the elliptic flow difference
(panels (a) and (b)) and ratio ((c) and (d)) of free neutrons vs. free protons produced in
semi-central $^{197}$Au+$^{197}$Au collisions at $E_{\rm lab}=400$~MeV$/$nucleon as
a function of the normalized transverse velocity $u_{t0}$
for two rapidity windows $|y_0|<0.4$ and $0.4<|y_0|<0.6$. }
\end{figure}

In order to enlarge the symmetry energy effect and eliminate model-parameter dependences and
other uncertainties of the individual flow observables, it is of interest to study the
neutron-to-proton elliptic flow difference $v_{2}^{n}$-$v_{2}^{p}$ and
ratio $v_{2}^{n}$/$v_{2}^{p}$~\cite{Cozma:2013sja,Guo:2013fka}.
In Fig.~\ref{fig3}, calculations with Skz4, SV-sym34, and SkI1 for $v_{2}^{n}$-$v_{2}^{p}$
(upper two panels) and $v_{2}^{n}$/$v_{2}^{p}$ (lower panels) are shown in two rapidity
windows: $|y_0|<0.4$ and $0.4<|y_0|<0.6$.
The $L$ value obtained with SV-sym34 is 80.95 MeV and lies between the two extremes
obtained with Skz4 and SkI1 (Table~\ref{skyrme}). Correspondingly, the flow differences and
ratios calculated with this force are also centered in all plots.
Furthermore, since the elliptic flows of neutrons and protons are negative within current
conditions, the $v_{2}^{n}$-$v_{2}^{p}$ differences decrease while the $v_{2}^{n}$/$v_{2}^{p}$
ratios increase with increasing $L$. With increasing $u_{t0}$, the sensitivity to the
symmetry energy of $v_{2}^{n}$-$v_{2}^{p}$ ($v_{2}^{n}$/$v_{2}^{p}$) is enlarged (reduced),
just because the absolute value of $v_2$ increases (Fig.~\ref{fig2}).
Finally, in both rapidity windows, but more so for $0.4<|y_0|<0.6$,
the isospin effect on $v_{2}^{n}$-$v_{2}^{p}$ and $v_{2}^{n}$/$v_{2}^{p}$ at
about $0.5<u_{t0}<1.5$ is visibly large. Provided that the statistical errors will be small
enough, intermediate rapidity windows may thus serve as promising kinematic regions in
future experiments aiming at extracting a more precise information on the density dependence
of the symmetry energy.

\begin{figure}[htbp]
\centering
\includegraphics[angle=0,width=0.9\textwidth]{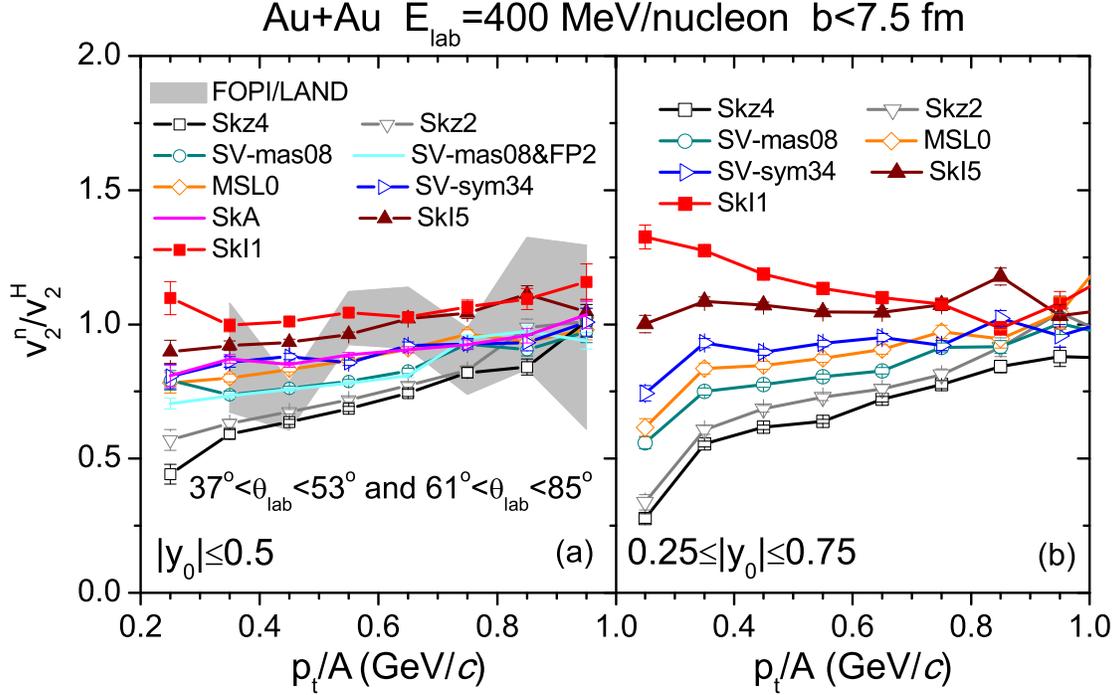}
\caption{\label{fig4}(Color online)
(a) Elliptic flow ratio of neutrons vs. hydrogen isotopes ($Z=1$) as a function of the transverse
momentum $p_t/A$, calculated with the indicated 9 Skyrme forces for central ($b<7.5$~fm)
$^{197}$Au+$^{197}$Au collisions at $E_{\rm lab}=400$~MeV/nucleon in the mid-rapidity
interval $|y_0|\leq0.5$ in comparison with the FOPI/LAND data (shaded area) reported in
Ref.~\protect\cite{Russotto:2011hq}; \\
(b) the same quantity calculated with the indicated 7 Skyrme forces for the intermediate rapidity
interval $0.25\leq|y_0|\leq0.75$. }
\end{figure}

In the following, we will re-compare existing FOPI/LAND data,
as reported in~\cite{Russotto:2011hq},
with calculated elliptic flow ratios of neutrons vs. protons or hydrogen isotopes.
For a better statistical accuracy in both calculations and experimental data, the elliptic
flow ratio of neutrons vs. hydrogen isotopes ($v_{2}^{n}$/$v_{2}^{H}$) is first considered as
it is known to be equally sensitive to the symmetry energy, especially at smaller transverse momenta
(cf. Fig.~\ref{fig3}) where the experimental errors are relatively small.
Figure~\ref{fig4} shows the comparison of the measured and the calculated
ratios $v_{2}^{n}$/$v_{2}^{H}$ as a function of the transverse momentum per nucleon $p_t/A$
($p_t/A = u_{t0} \cdot 0.431$~GeV/c at $E_{\rm lab}=400$~MeV/nucleon and $A$ is the mass number
of the emitted particles). SV-mas08\&FP2 denotes the result calculated with SV-mas08 and
the FP2 parameterization of the NNECS~\cite{Wang:2013wca,Li:2011zzp}. All calculations were
performed for the indicated impact-parameter and rapidity intervals and gated with the range
of laboratory angles accepted by LAND.

Similar to the results shown in Fig.~\ref{fig3}, the $v_{2}^{n}$/$v_{2}^{H}$ ratio increases with
increasing $L$, from Skz4 to SkI5, and its spreading steadily grows when moving to the low
transverse momentum region. When interpreting the results in more detail, it is seen that
results calculated with SV-sym34 and SkA, for which the difference in $L$ is only about 6 MeV,
are almost overlapped even though the difference in $K_0$ is as large as almost 30 MeV and that
of the effective mass ratio $m^*/m$ is 0.3. It illustrates  the sensitivity of the elliptic
flow ratio to the stiffness of the symmetry energy and not to the incompressibility of the
nuclear EoS.
Also the big difference in $m^*/m$ does not affect much the
$v_{2}^{n}$/$v_{2}^{H}$ ratio, which, at first glance, is not in agreement with previous work
addressing the contribution of the momentum dependent term to flow observables (see, e.g.,
Refs.~\cite{Cozma:2013sja,Hartnack:1994zz}). However, as discussed for the $\rho\tau$ term in
Eq.~\ref{urho}, the momentum dependent contribution from the Skyrme potential energy density
functional is very limited for the colliding conditions considered in the current work.
The momentum dependence comes mainly from the optical potential which originates to a large extent
from a type of Lorentz force in a relativistic description. Actually, another parameterization
of the optical potential suggested in Ref.~\cite{Hartnack:1994zz} has also been tested and no
visible effect on the elliptic flow ratio has been found, in line with the result of
Ref.~\cite{Zhang:2012fc}.

Furthermore,
the results obtained with SV-mas08\&FP2 and SV-mas08 (i.e. with FP4) track each other closely,
confirming the weak effect of the in-medium NNECS on the elliptic flow ratio already observed in
Ref.~\cite{Russotto:2011hq}. Note that the isospin dependence in both, the FP2 and FP4 NNECS
parameterizations is not changed which still represents an open question. Accordingly, the relative
contributions of momentum and density modified NNECS to the flows of neutrons and hydrogen isotopes
should not change largely. Certainly, a larger uncertainty assumed in the medium modifications of
NNECS will definitely change the flow ratio visibly due to the change in the total collision
numbers. In that case, however, the individual elliptic flows of neutrons and hydrogen isotopes will
also vary largely, deviating from the experimental data shown in Fig.~\ref{fig2} and thus from
our desired precondition. We may, therefore, conclude that the systematically increasing values
of the elliptic flow ratio $v_{2}^{n}$/$v_{2}^{H}$ calculated with the selected forces are
mainly due to the increase of the stiffness of the symmetry
energy and not to other changes of isoscalar components of the dynamic transport.

\begin{figure}[htbp]
\centering
\includegraphics[angle=0,width=0.7\textwidth]{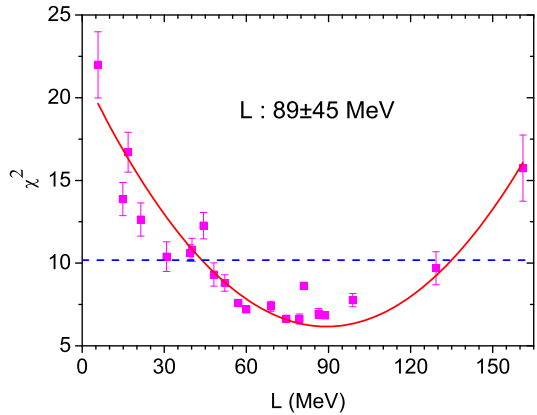}
\caption{\label{fig5}(Color online)
The total $\chi^2$ characterizing the fit results obtained with the 21 studied
Skyrme forces as a function of the slope parameter $L$. The smooth curve is a quadratic
fit to the total $\chi^2$, and the horizontal dashed line is used to determine the error of $L$
within a 2-$\sigma$ uncertainty.}
\end{figure}

Figure~\ref{fig4} (b) shows the calculation results for the intermediate rapidity
window $0.25\leq|y_0|\leq0.75$, for the same impact parameter and rapidity interval but without
the gate on laboratory angles.
It is clearly seen that the differences of the various predictions steadily grow as one
moves to the region of low transverse momentum.
The $v_{2}^{n}$/$v_{2}^{H}$ ratio in the rapidity window $0.25\leq|y_0|\leq0.75$ seems considerably
more sensitive to the density dependent symmetry energy than in the mid-rapidity interval
$|y_0|\leq0.5$, thus offering interesting opportunities for future experiments.

The results of fitting the transverse-momentum dependent flow ratio (Fig.~\ref{fig4} (a))
are shown in Fig.~\ref{fig5} as a function of the slope parameter $L$.
The symbols represent the total $\chi^2$ as calculated with the 21 Skyrme interactions.
The variation with $L$ is well described with a quadratic fit with an adjusted coefficient
of determination (Adj. R-square) of 0.89. The obtained minimum of $\chi^2$ corresponds to a
slope parameter value  $L=89\pm45$~MeV within a 2-$\sigma$ uncertainty.
The so obtained constraint is rather close to the $L=83\pm52$~MeV (2-$\sigma$ uncertainty)
obtained previously~\cite{Russotto:2011hq}. The larger error in the latter case incorporates
estimates of systematic deviations observed when studying the impact parameter dependence
of the model comparisons. In fact, the fit result of $\gamma = 0.98 \pm 0.35$ obtained by
Russotto et al. with the FP2 parameterization, corresponding to $L=89\pm46$~MeV (2 $\sigma$),
is identical to the value favored here, again supporting the sensitivity to the density
dependence of the symmetry energy while details of the method are of reduced importance.

\begin{figure}[htbp]
\centering
\includegraphics[angle=0,width=0.9\textwidth]{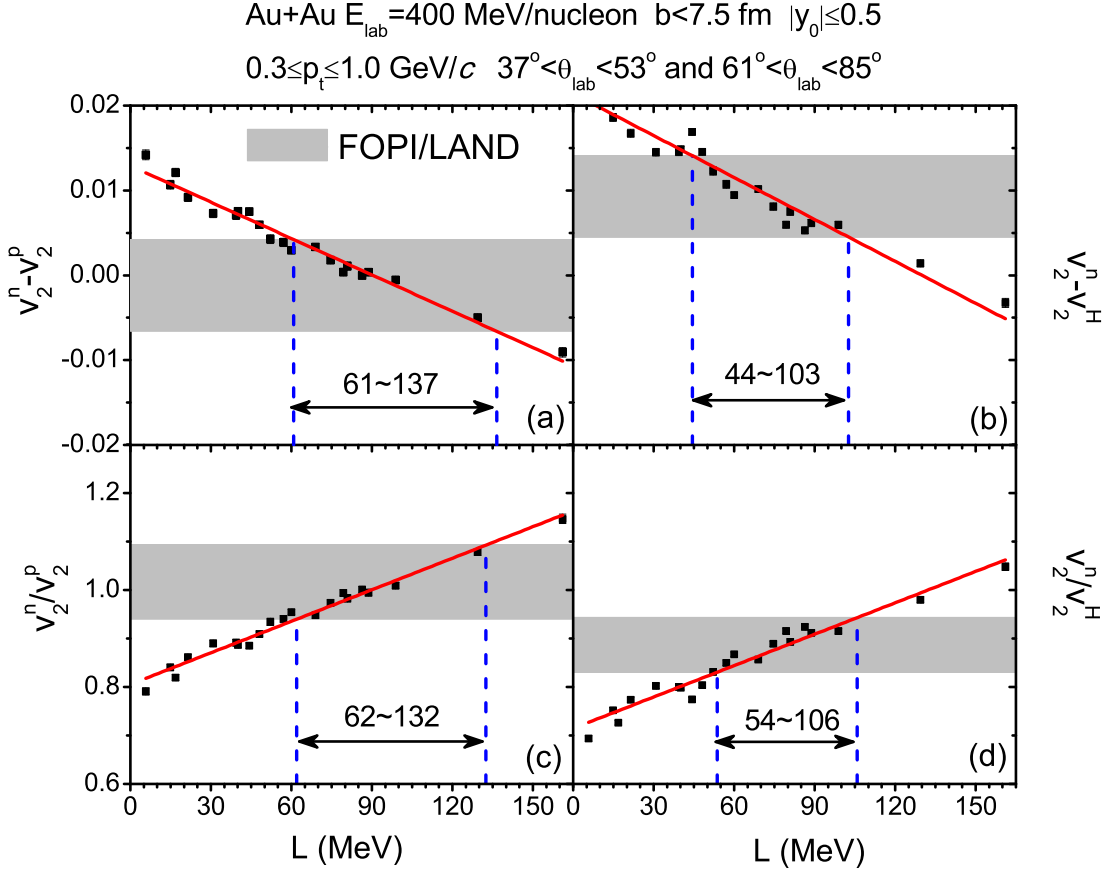}
\caption{\label{fig6}(Color online) The elliptic flow differences $v_2^{n}-v_2^{p}$ (a) and
$v_2^{n}-v_2^{H}$ (b) and the elliptic flow ratios $v_2^{n}/v_2^{p}$ (c) and $v_2^{n}/v_2^{H}$ (d)
produced in moderately central ($b<7.5$~fm) $^{197}$Au+$^{197}$Au collisions at
$E_{\rm lab}=400$~MeV/nucleon are shown as a function of the slope parameter $L$. In each plot,
the gray shaded region indicates the $p_t/A$-integrated experimental data \cite{Russotto:2011hq},
full squares denote UrQMD calculations with the studied set of Skyrme forces, while the lines
represent linear fits to the calculations.}
\end{figure}

It is interesting to see that the present result is slightly higher but still overlaps well
with recent constraints for the symmetry energy at subnormal and normal densities, and even
with constraints from astrophysical observations (see, e.g.,
Refs.~\cite{Li:2012mw,Chen:2012pk,Tsang:2012se,Lattimer:2012xj,steiner:2013}). Differences may arise to the extent
that higher densities are probed with the present flow observables.
We also note that a larger $L$ reduces 
the difficulty of parameterizing the symmetry energy discussed by Dong et al. since $K_{\rm sym}$
is close to zero near the present central $L \approx 90$~MeV, according to the correlation
established there~\cite{Dong:2012zza}. Among the Skyrme forces in the favored $L$ interval,
we find the MSL0 parameter set ($L = 60$~MeV) which was based on a series of analyses of the
neutron-skin thickness of heavy nuclei,
isospin diffusion, and the double neutron/proton ratio in HICs at intermediate energies~\cite{Chen:2010qx}. This force, as well as members of the
families of SV-sym34 ($L = 81$~MeV) and Ska35s25 ($L = 99$~MeV), belong to the highly selected
CSkP set of Skyrme forces favored by the analysis of Dutra et al.~\cite{Dutra:2012mb}.

The quadratic behavior of $\chi^2$ indicates that the flow ratio correlates linearly with $L$
among the set of Skyrme forces studied. This is illustrated in Fig.~\ref{fig6} for the four
observables $v_2^{n}-v_2^{p}$, $v_2^{n}-v_2^{H}$, $v_2^{n}/v_2^{p}$, and $v_2^{n}/v_2^{H}$. A linear
fit describes the calculated correlation rather well in all cases. A comparison is made with
the FOPI-LAND results obtained for these observables when sorted into the transverse-momentum
interval 0.3~GeV/c$\le p_t \le$ 1.0~GeV/c. The selected intervals of $L$ (with 1-$\sigma$
uncertainty) largely overlap but also exhibit differences of up to about 25 MeV of their
central values whose numerical average is close to $L=85$~MeV.
It will be interesting to see more precise data
becoming available as, e.g., from the recent ASY-EOS (S394) experiment at GSI \cite{Russotto:2012jb},
which will be important also for testing the consistency of the model predictions.

As reported in Ref.~\cite{Cozma:2013sja}, a slope parameter of $L=122\pm57$~MeV was extracted with
the help of the T\"{u}bingen QMD model. This is stiffer by about 30 MeV than the typical values
that have been obtained here. As it turns out, this is only partly due to a possible model dependence
because numerical uncertainties arise also from how the data are sorted. Cozma et al. start from an
impact parameter dependent set of experimental values obtained by using the ratio ERAT of total
transverse-to-longitudinal kinetic energies in the center-of-mass system~\cite{FOPI:2011aa}
for sorting according to centrality. Weighted averages were used to represent the impact-parameter
integrated ($b \le 7.5$~fm) flow values. Even though the results are fully consistent
within errors, they deviate slightly numerically from the multiplicity sorted values used
by Russotto et al. as well as here.
The numerical average deduced from comparing the neutron/proton and neutron/hydrogen
differences and ratios with the calculations shown in Fig.~\ref{fig6}
is near $L=110$~MeV, reflecting a 25-MeV systematic uncertainty due to differences in
impact-parameter sorting and in fitting correspondingly different azimuthal angular distributions.
The remaining difference to $L=122$~MeV may be ascribed to a residual model dependence of the
UrQMD and T\"{u}bingen-QMD analyses. It is small in comparison to the quoted uncertainty
$|\Delta L| = 57$~MeV which, to a large part, arises from unknown properties of the
ingredients of transport models, including their isoscalar sector~\cite{Cozma:2013sja}.

To summarize, the recently released FOPI experimental data of the transverse-velocity dependence of
elliptic flow of protons has been well described with the updated UrQMD transport model
in which the Skyrme potential-energy-density functional is adopted for the mean-field part.
The transverse-momentum dependent elliptic flow ratios $v_{2}^{n}/v_{2}^{p}$ and differences
$v_{2}^{n}$-$v_{2}^{p}$ were shown to exhibit a larger sensitivity to the stiffness of the symmetry
energy in the intermediate rapidity intervals $0.4<|y_0|<0.6$ than at mid-rapidity. As a function of
the transverse velocity, the sensitivity of the flow ratios is enhanced at the lower and that of
the differences at the higher transverse velocities. By comparing the calculations with the
transverse-momentum dependent FOPI/LAND flow ratio $v_{2}^{n}/v_{2}^{H}$ the slope parameter
of the density-dependent symmetry energy is extracted to be $L = 89 \pm 45$~MeV within a 2-$\sigma$
confidence limit. Linear correlations of $L$ with the predictions for
ratios and differences for neutrons vs. protons and neutrons vs. all hydrogen
isotopes have been established for the selected class of phenomenological forces.
The results obtained from the comparison with the four transverse-momentum
integrated experimental values are spread over an interval of $\Delta L \approx 25$~MeV
but found to be consistent with each other within errors. Their numerical average is
$L \approx 85$~MeV. A slightly larger slope parameter $L \approx 110$~MeV is obtained from a
similar comparison with the flow data published in Ref.~\cite{Cozma:2013sja} for which the
transverse-vs.-longitudinal kinetic energy ratio ERAT has been used for the impact-parameter
determination instead of the particle multiplicity used in Ref.~\cite{Russotto:2012jb}.
More precise data will be needed to clarify to what extent these differences indicate residual
systematic uncertainties or, possibly, mainly reflect the general statistical limits of the
presently available experimental flow data.

The presented results are in full agreement with the previous studies performed with the UrQMD or
T\"{u}bingen QMD models, indicating a moderately soft to linear density dependence of the symmetry
energy. They contrast with diverging results obtained from the comparisons of IBUU or LQMD model
calculations with the FOPI $\pi^-/\pi^+$ ratios from which both, extremely soft and extremely stiff
behaviors were extracted. The obtained slope parameter of $L \approx 90 - 110$~MeV is larger than
the typical $L = 60$~MeV resulting from nuclear structure studies and reactions at lower energies.
If confirmed with more precise data, this may reflect the higher densities probed at the present
energy as well as the approximations made with parameterizations and extrapolations.
It is clear that new experiments addressing the strength and density dependence of the symmetry
energy, as presently performed or in preparation at several laboratories, will play an important role
in clarifying these questions.

On the theoretical side, the non-equilibrium effect is one of the most important characteristics
of heavy ion collisions at
intermediate energies for which the isospin degree of freedom can be used as a sensitive probe.
The issue of the degree of isospin equilibrium should receive more attention, including the
investigation of the isospin dependence in both, the momentum-modified mean-field potentials and
the binary scatterings representing the in-medium nuclear interaction. In a future study, this
problem will be addressed within the same microscopic transport model.

\begin{acknowledgments}
The authors thank W. Reisdorf for providing experimental data and acknowledge
support by the computing server C3S2 in Huzhou Teachers
College and the warm hospitality of the Frankfurt Institute for Advanced Studies (FIAS), Johann Wolfgang Goethe-Universit$\ddot{a}$t, Germany. The work is supported in part by the National Natural
Science Foundation of China (Nos. 11175074 and 11375062), the Qian-Jiang Talents Project of Zhejiang Province
(No. 2010R10102) and the project sponsored by SRF for ROCS, SEM.
\end{acknowledgments}

\end{document}